\documentclass[osajnl,twocolumn,showpacs,superscriptaddress,10pt]{revtex4-1} 

\usepackage{amsmath,amssymb,graphicx}
 \usepackage[french]{babel}
  \usepackage[T1]{fontenc}
 \usepackage[latin1]{inputenc}
 \usepackage{srcltx}
\begin{document}

\title{Heterodyne Holography with full control of both signal and reference arms
}

\author{Michel Gross}
\address{Laboratoire Charles Coulomb (L2C), UMR 5221 CNRS-Universit\'{e} de Montpellier, Montpellier, F-France}

\email{michel.gross@univ-montp2.fr}

\begin{abstract}
Heterodyne holography  is a variant of phase shifting holography in which reference and signal arms are controlled by acousto optic modulators. In this review paper, we will briefly describe  the method and its properties,  and  we will illustrate  its advantages in experimental applications.
\end{abstract}

\maketitle

Ocis: 090.1995: Holography: Digital holography;
040.2840: Detectors-Heterodyne.

\bigskip

\textbf{Citation:} Michel Gross, Heterodyne holography with full control of both the signal and reference arms,\textit{ Appl. Opt.} \textbf{55}, pA8--A16 (2016) doi: 10.1364/AO.55.0000A8
\url{https://www.osapublishing.org/ao/abstract.cfm?uri=ao-55-3-A8}

\section{Introduction}

Heterodyne holography  \cite{ le2000numerical,le2001synthetic} is a technique firstly introduced 15 years ago realized by modifying a phase shifting holography setup \cite{yamaguchi1997phase}. The main advantage of heterodyne holography is that the  frequency, phase and amplitude of both reference and signal arms are controlled by acousto optic modulators (AOM).
%
%
By shifting the frequency $\omega_{LO}$
of the reference  beam, called here local oscillator, with respect to the frequency $\omega_I$ of illumination, heterodyne holography
is  able to detect the light scattered by the object  at any  frequency $\omega$ close to $\omega_{LO}$, which can be equal or different than
the  frequency $\omega_I$ of illumination of the object.

If the light is scattered by the object over a broad continuous frequency spectrum, heterodyne holography must be combined with off-axis holography in order to separate the images corresponding to the +1 and -1 holographic grating orders.  Indeed, in that case,   the +1 and -1  holographic images are both non-zero, since they correspond to the detection of signals at frequencies that are different but  very close and
within the object broad frequency spectrum \cite{atlan2007accurate}.  This point will be discussed in section \ref{section_freq_response}.  Heterodyne holography in off axis configuration has been used since 2003 \cite{gross2003shot}.

%
%

Heterodyne holography performs heterodyne detection with a slow 2D multi pixels detector  (CCD or  CMOS camera). The  bandwidth BW of the detection is narrow (10 to 1000 Hz depending on the camera), but the beat frequency  $\omega-\omega_{LO}$ can be large (up to several MHz  \cite{gross2003shot}). Heterodyne holography  must be distinguished from optical scanning holography that uses a fast mono pixel detector (photodiode), and that makes mechanical scans along the $x$ and $y$ axis  to get  holograms \cite{Poon2009optical}.

We will briefly describe the heterodyne holographic technique   and illustrate  its advantages on experimental examples.


\section{Principle and properties of heterodyne holography  }

\subsection{Typical setup }

\begin{figure}[h]
\begin{center}
  \includegraphics[width=6.5 cm]{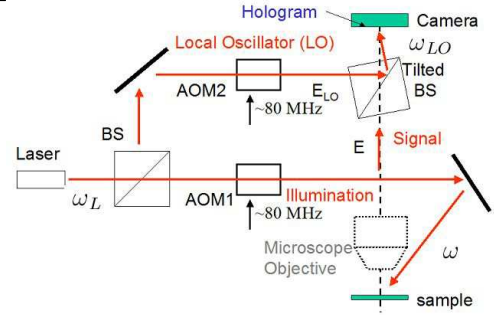}\\
  \caption{Typical heterodyne holography setup in transmission geometry.  BS: beam splitter;
AOM1, AOM2: acousto-optic modulators;    $E_{LO}$ and $E$:  reference (i.e. local oscillator LO) and object
fields whose optical frequencies are $\omega_{LO}$ and $\omega$; $\omega_{AOM1/2}$: driving frequencies  ($\simeq$ 80 MHz) of the acousto optics modulators AOM1 and
AOM2.}\label{Fig__setup}
\end{center}
\end{figure}

Figure \ref{Fig__setup} shows a typical heterodyne holography setup. The frequency and the amplitude  of both the illumination and the reference arms are  controlled by the two acousto optics modulator (Bragg cells) AOM1 and AOM2. AOM1, which  is driven   by a  Radio Frequency  (RF) signals at $\omega_{AOM1}$  (with $\omega_{AOM1} \simeq 80 $ MHz),  shifts  the   frequency $\omega_I$ of illumination so that
\begin{eqnarray}\label{Eq_omega_I}
 \omega_I&=&\omega_L+ \omega_{AOM1}
\end{eqnarray}
where $\omega_L$ is the optical frequency of the main laser. Similarly AOM2 shift the frequency $\omega_{LO}$ of the local oscillator arm so that
\begin{eqnarray}\label{Eq_omega_LO}
 \omega_{LO}&=&\omega_L+ \omega_{AOM2}
\end{eqnarray}
An angularly tilted beam splitter BS  mixes the signal field scattered by the sample ${\cal E}(t)= E e^{j\omega t} $  and reference field ${\cal E}_{LO}(t)= E_{LO} e^{j\omega_{LO} t}$. Thus, the interference pattern $I$ that is recorded by the CCD camera is:
\begin{eqnarray}
  I (t)&=& |{\cal E}(t)+ {\cal E}_{LO}(t)|^2\\
\nonumber   &=&|E|^2 + | E_{LO}|^2+ E E_{LO}^*~ e^{-j (\omega_{LO} -\omega)t} +\textrm{c.c.}
\end{eqnarray}
where $\textrm{c.c.}$ is the complex conjugate of  the $E E_{LO}^*$ term i.e. ~  $\textrm{c.c. }=E^* E_{LO}~ e^{+j (\omega_{LO} -\omega)t}$.
Sequence of frames $I_n$ are recorded by the camera, with:
\begin{eqnarray}\label{Eq_sinc_time}
  I_n&=& \frac{1}{T}  \int_{t=nT_{CCD}-T/2}^{nT_{CCD}+T/2}  dt  |{\cal E}(t)+ {\cal E}_{LO}(t)|^2\\
\nonumber   &=&|E|^2 + | E_{LO}|^2\\
\nonumber&&~~~  + E E_{LO}^* \left[\frac{1}{T}\int_{t=nT_{CCD}-T/2}^{nT_{CCD}+T/2}  dt ~ e^{-j (\omega_{LO} -\omega)t}\right] + \textrm{c.c.}\\
\nonumber   &=&|E|^2 + | E_{LO}|^2 \\
\nonumber&&~~~  + E E_{LO}^* ~\textrm{sinc}\left( (\omega-\omega_{LO})T/2 \right)~ e^{-j (\omega_{LO} -\omega)t_n} + \textrm{c.c.}
\end{eqnarray}
where  $t_n= n T_{CCD}$ with  $n$  integer, $T_{\textrm{CCD}}=2\pi/\omega_{CCD}$, and   $\omega_{CCD}/2\pi$ is the camera frame frequency. The  sinc factor of Eq. \ref{Eq_sinc_time} corresponds to the integration
of the beat signal of frequency  $\omega_{LO} -\omega$  over
the camera exposure time $T$.

Note that a  microscope objective can added to the setup to perform heterodyne holographic microscopy
\cite{warnasooriya2010imaging,verpillat2011dark,verrier2014laser}.

\subsection{Different number of frames detection}

\begin{figure}[h]
\begin{center}
  \includegraphics[width=5 cm]{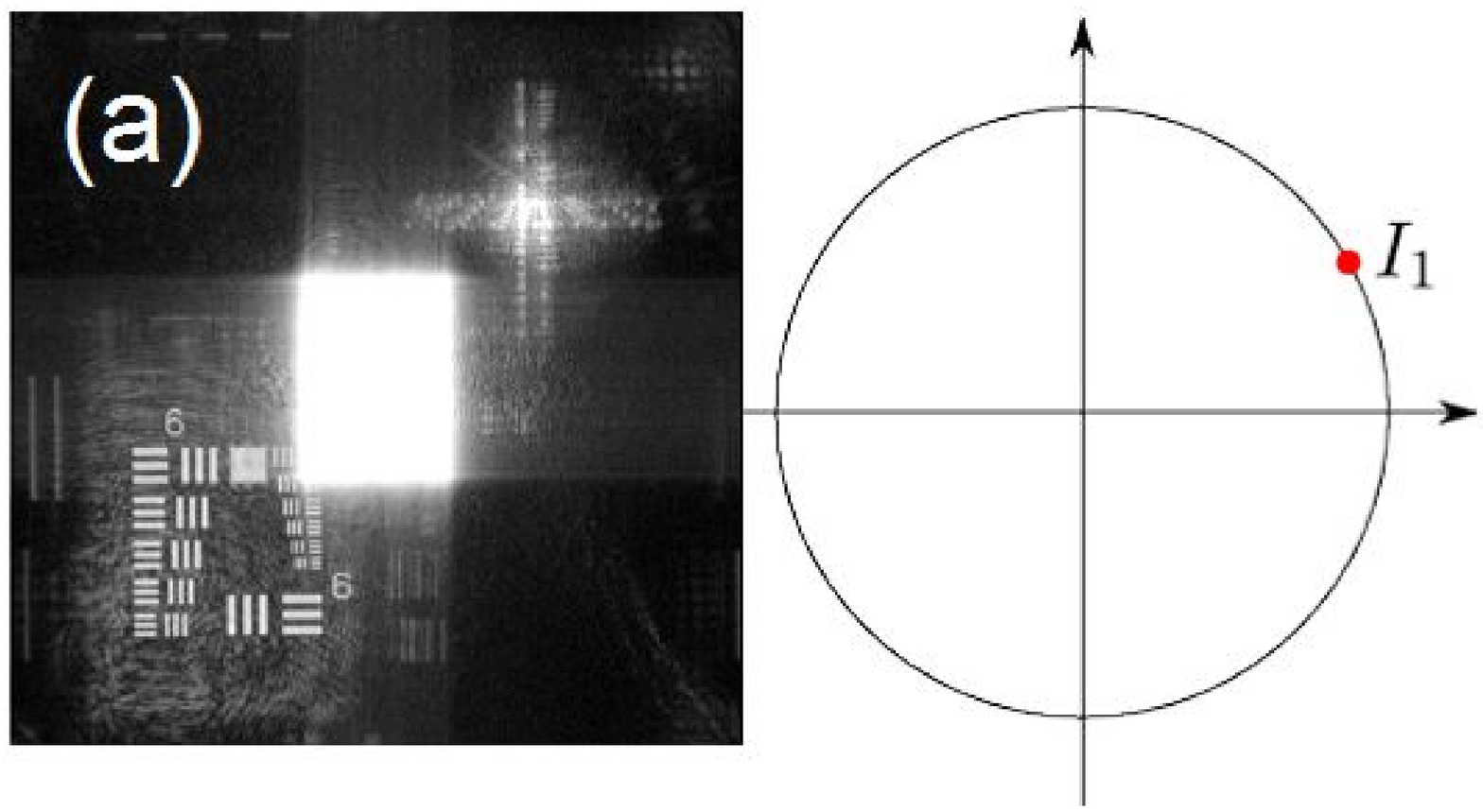}\\
  \includegraphics[width=5 cm]{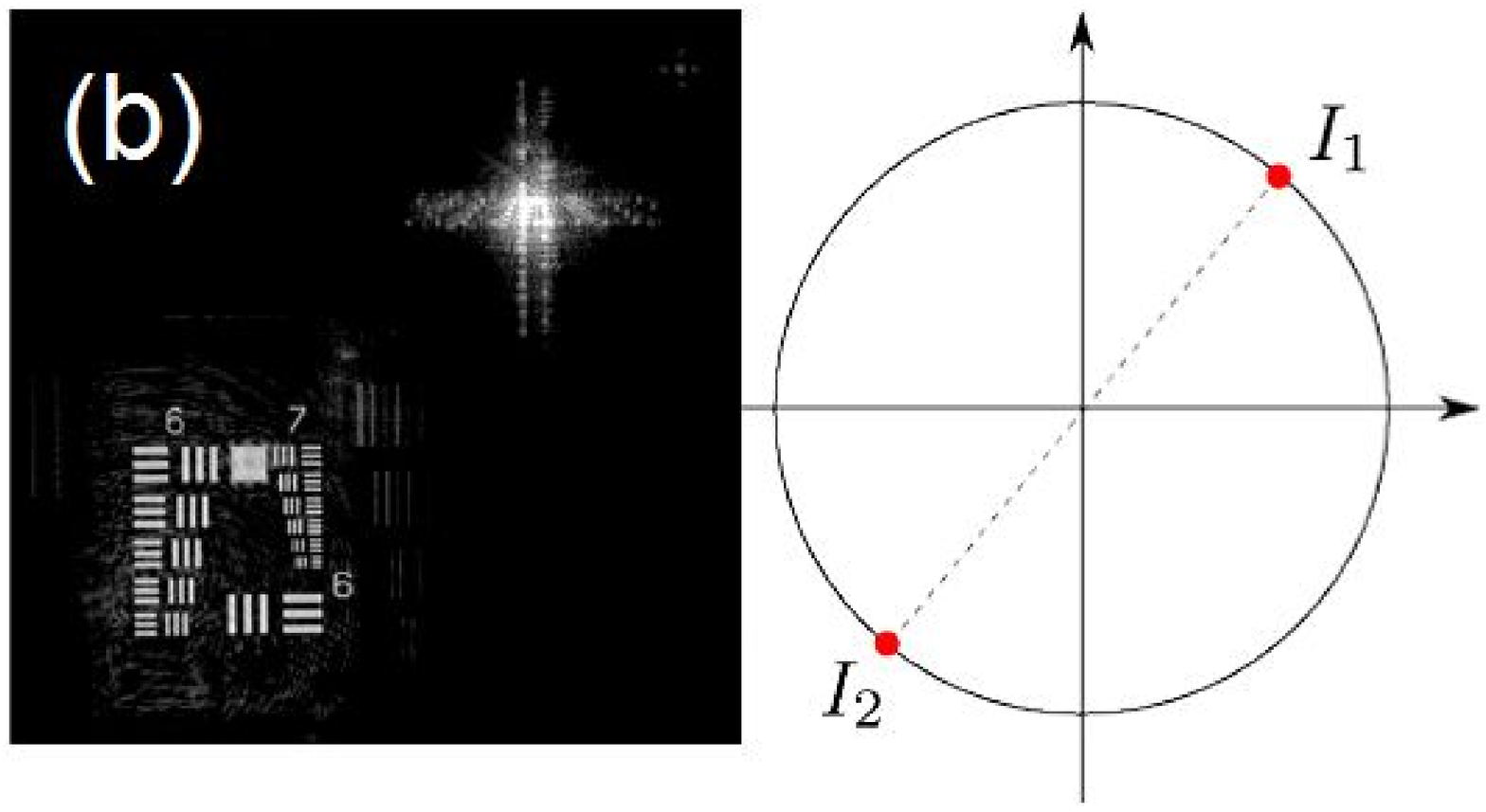}\\
  \includegraphics[width=5 cm]{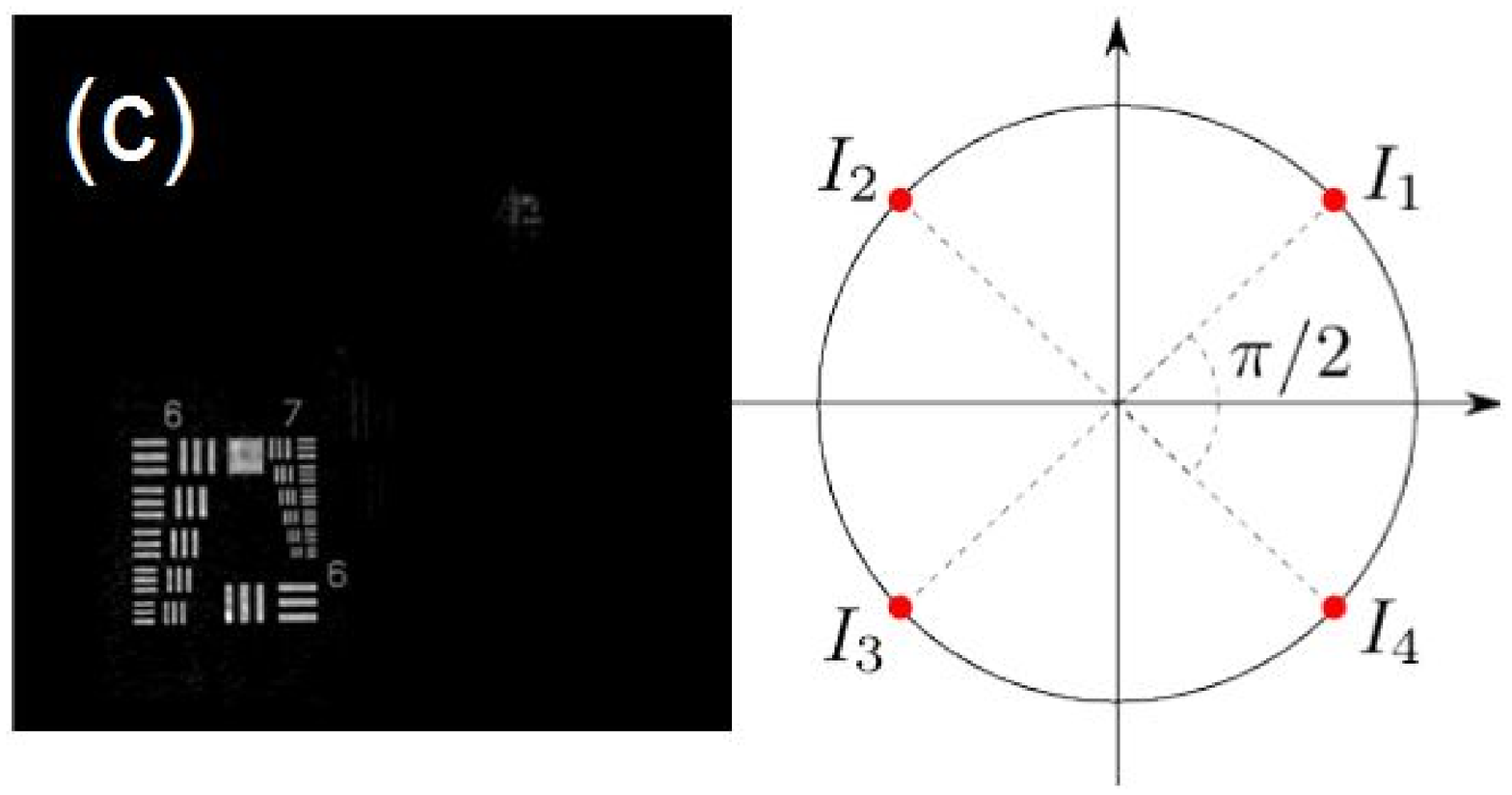}\\
  \caption{(Left hand side) Reconstructed images of a USAF target obtained with 1 frame (a), 2 frames (b) and 4 frames (c) heterodyne  holography. (Right  hand side) Phase of the signal field $E$ (relative to $E_{LO}$) for the successives frames $I_1$, $I_2$ ... of the sequence.}\label{Fig__124phase}
\end{center}
\end{figure}

Different types of holographic detection schemes can be implemented with the setup of Fig. \ref{Fig__setup} .
To illustrate this point, we have considered a USAF target sample.
Since the target does not move, the light is scattered by the target at the illumination  frequency $\omega= \omega_I$.

%
%
%
%
%

\subsubsection{One frame off axis holography}

One frame off axis holography  is  made by choosing  $\omega_{AOM2} =\omega_{AOM1}$ (so that $\omega_{LO}=\omega_I=\omega$), and   $H=I_1$ for the hologram. We get then:
\begin{eqnarray}\label{Eq_H_one_phase}
  H  &=&I_1=|E|^2 + | E_{LO}|^2 \\
\nonumber  &&~~~~~~~~~~+ \textrm{sinc}((\omega-\omega_{LO}) T/2) (E E_{LO}^* + E^* E_{LO} )
\end{eqnarray}
Figure \ref{Fig__124phase} (a) shows the reconstructed image  obtained in that case. The image exhibits 3 bright zones.
The bright square in the center of the reconstructed image is the zero grating order. It corresponds to the $|E_{LO}|^2 + |E|^2$ terms.
The blurred bright zone in the upper right side of the image is the $-1$ grating order. It corresponds to  $E_{LO}E^*$.
Lastly, the USAF image that is sharp  in the lower left  side of the reconstructed image is the
$+1$ grating  order that corresponds to $E_{LO}^* E$.

\subsubsection{Two frames  phase shifting holography}

Phases shifting holography with 2 frames is made by tuning the local oscillator frequency  to have  $\omega_{AOM2} - \omega_{AOM1} = \omega_{CCD}/2$ (so that $\omega_{LO}-\omega_I =  \omega_{CCD}/2$). For detection at the illumination frequency (i.e. for $\omega=\omega_I)$, the phase factor  $e^{-j (\omega_{LO} -\omega)t_n}$ of Eq. \ref{Eq_sinc_time} becomes thus equal to $-1^{n-1}$. We  get for $I_1$ and $I_2$:
\begin{eqnarray}\label{Eq_I_1_I_2_2_phases}
    I_1 &=& |E|^2 + | E_{LO}|^2 \\
\nonumber     &&~~~~~~~~~~~+ \textrm{sinc}((\omega-\omega_{LO}) T/2) (+ E E_{LO}^* + E^* E_{LO} )\\
\nonumber   I_2 &=& |E|^2 + | E_{LO}|^2 \\
\nonumber     &&~~~~~~~~~~~+ \textrm{sinc}((\omega-\omega_{LO}) T/2) (- E E_{LO}^* - E^* E_{LO} )
\end{eqnarray}
By choosing $H=I_1-I_2$, we get:
\begin{eqnarray}\label{Eq_H_2_phases}
  H  &=& I_1-I_2\\
 \nonumber  &=&2~\textrm{sinc}( (\omega-\omega_{LO} T/2)) (  E E_{LO}^* +  E^* E_{LO} )
\end{eqnarray}

Figure \ref{Fig__124phase} (b) shows the  2 frames reconstructed  image. This image exhibits only 2 bright zones,  which correspond to the orders $+1$ and $-1$ i.e. to $E E_{LO}^*$ and $E^* E_{LO}$.

\subsubsection{Four frames  phase shifting holography }

Four phases phase shifting holography is made with  $\omega_{AOM2} - \omega_{AOM1} = \omega_{CCD}/4$ so that   $\omega_{LO}=\omega_I   + \omega_{CCD}/4$. For $t=t_n$, the phase factor  $e^{-j (\omega_{LO} -\omega)t}$ is thus equal to $(-j)^{n-1}$. We  get:
\begin{eqnarray}\label{Eq_I_1_I_4_4_phases}
    I_1 &=& |E|^2 + | E_{LO}|^2 \\
\nonumber     &&~~+\textrm{sinc}( (\omega-\omega_{LO} T/2)) ( E E_{LO}^* + E^* E_{LO})\\
\nonumber   I_2 &=& |E|^2 +| E_{LO}|^2\\
\nonumber     &&~~ + \textrm{sinc}( (\omega-\omega_{LO} T/2)) (-j E E_{LO}^* +j E^* E_{LO})\\
\nonumber   I_3 &=& |E|^2 +| E_{LO}|^2\\
\nonumber     &&~~+ \textrm{sinc}( (\omega-\omega_{LO} T/2))( - E E_{LO}^* - E^* E_{LO})\\
\nonumber   I_4 &=& |E|^2+| E_{LO}|^2\\
 \nonumber     &&~~ +\textrm{sinc}( (\omega-\omega_{LO} T/2))( +j E E_{LO}^* -j E^* E_{LO})
\end{eqnarray}
By choosing $H=(I_1-I_3)+j(I_2-I_4)$, we get:
\begin{eqnarray}\label{Eq_H_4_phases}
  H  &=& (I_1-I_3)+j(I_2-I_4)\\
 \nonumber  &=& 4 ~\textrm{sinc}( (\omega-\omega_{LO} T/2)) E E_{LO}^*
\end{eqnarray}
Figure \ref{Fig__124phase} (c) shows the  4 frames reconstructed  image. This image exhibits only the +1 grating order   $E E_{LO}^*$, which gives  a sharp image of the USAF  target.

\subsection{Frequency response $\eta(\omega)$ of the holographic heterodyne detection}\label{section_freq_response}

\begin{figure}[h]
\begin{center}
  \includegraphics[width =5 cm]{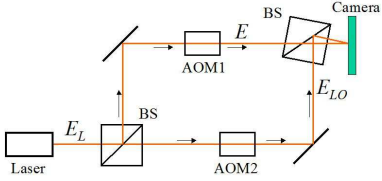}
  \caption{Setup of the holographic experiment used to measure the frequency response $\eta(\omega)$. BS: beam splitter;
AOM1, AOM2: acousto-optic modulators;    $E_{LO}$ and $E$:  reference (i.e. local oscillator LO) and object
fields whose optical frequencies are $\omega_{LO}$ and $\omega$}\label{Fig_Fig_test_setup}
\end{center}
\end{figure}

In the USAF target experiment, the position of the sample do not change and the light is scattered  at the illumination  frequency   $\omega= \omega_I$. In some other cases, the sample could be  moved and the frequency of the scattered light can be Doppler shifted. Thus, it is important to know the frequency response $\eta(\omega)$ of the holographic heterodyne detection.

\begin{figure}[h]
\begin{center}
  \includegraphics[width =7.5 cm]{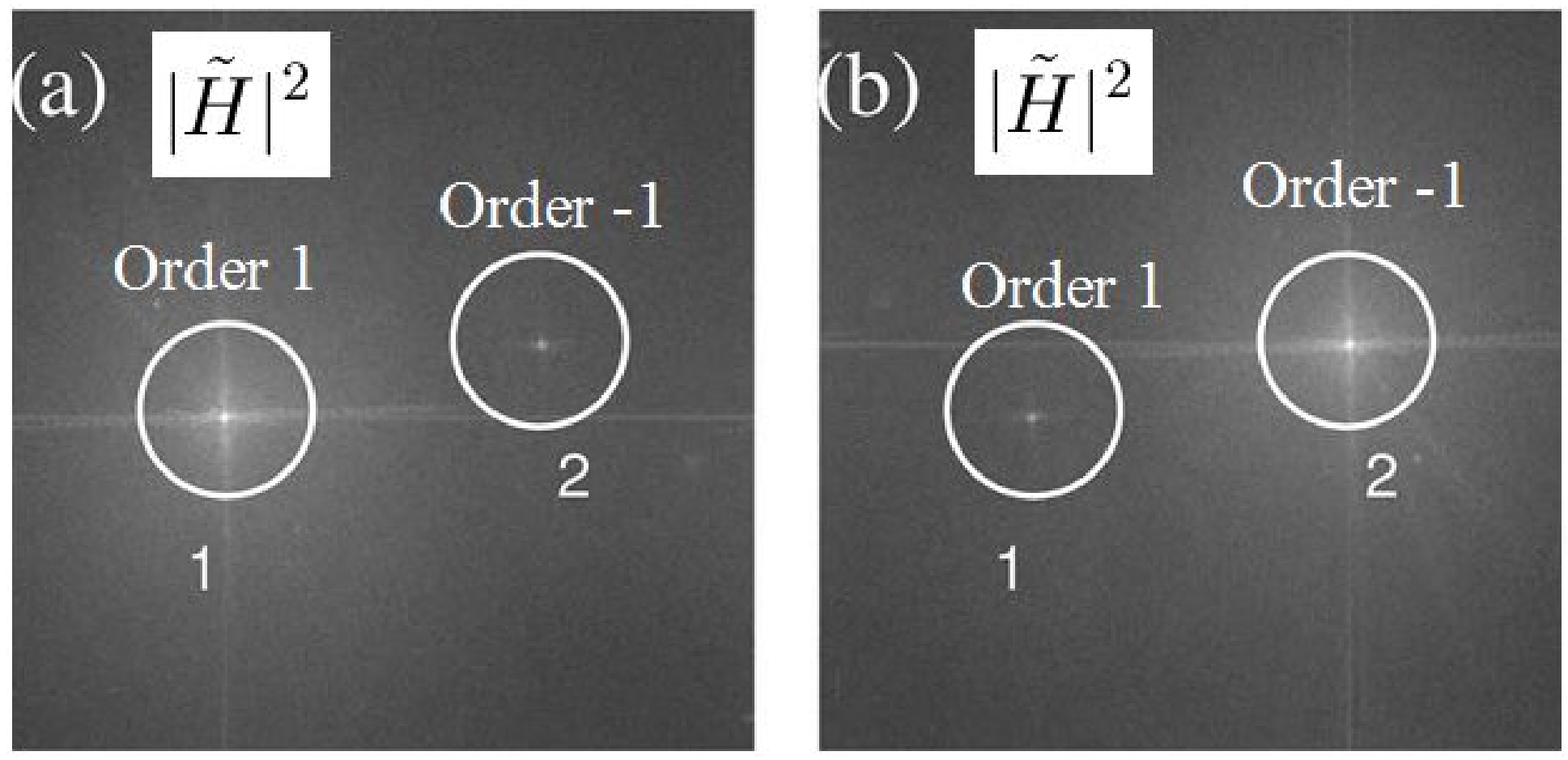}
  \includegraphics[width = 7.5 cm]{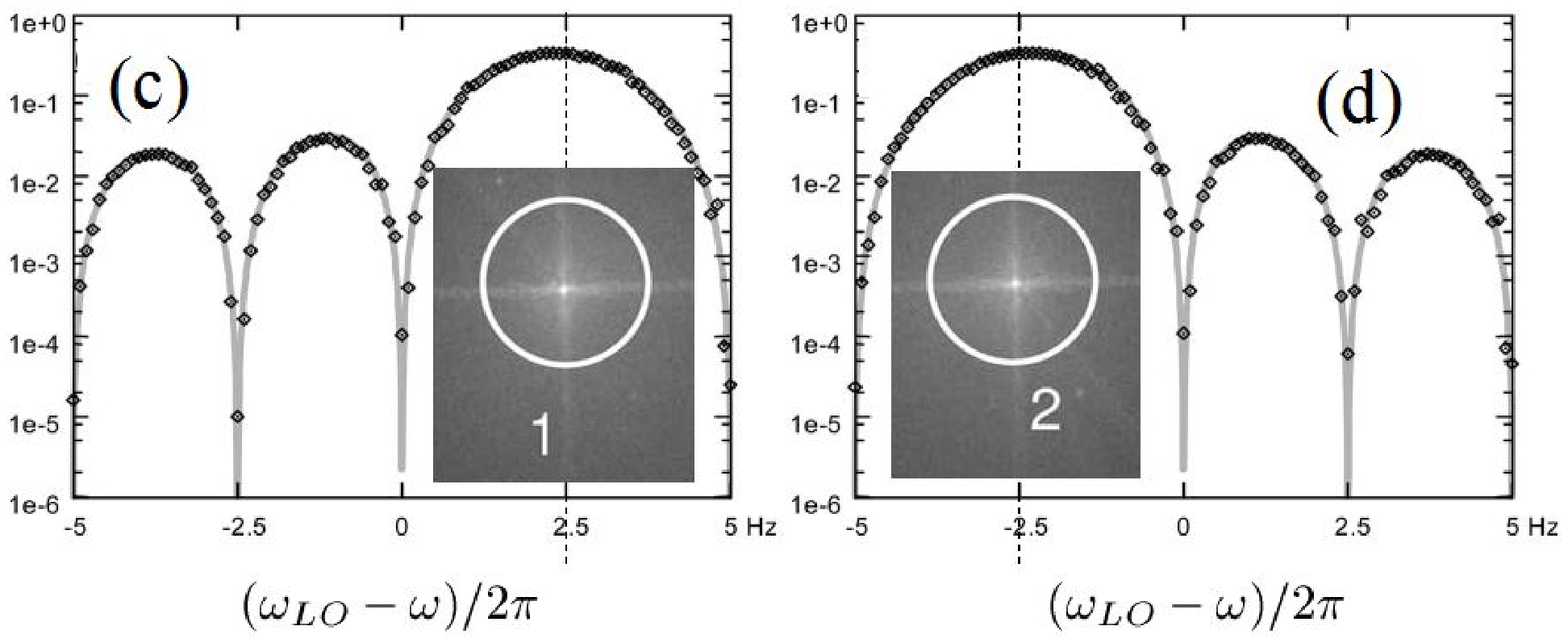}\\
  \caption{ (a,b) Fourier space holograms $|\tilde H|^2$ obtained with $\omega_{AOM2}-\omega_{AOM1} = + \omega_{CCD}/4$ (a) and $- \omega_{CCD}/4$ (b) with $\omega_{CCD}/2\pi= 10$ Hz.  Weights $W_{\pm1}$ of the signal $|H|^2$ summed over regions 1 and 2 of Fig 4 (a) and (b)
are plotted in (c,d). $W_{+1}$ is plotted in (c), $W_{-1}$ in (d). Horizontal axis is ($\omega_{LO}-\omega)/2\pi$. Points are measurements: $W_{\pm 1}$  given by
Eq. 11, solid gray line is theory: $|\eta|^2$ given by Eq. 13 \cite{atlan2007accurate}.}\label{Fig_Fig_accurate_phase}
%
%
%
\end{center}
\end{figure}

To determine $\eta$  an holographic control experiment illustrated by Fig. \ref{Fig_Fig_test_setup} has been performed \cite{atlan2007accurate}. The experiment is made by recording sequence of frames $I_n$ for  different frequency $\omega=\omega_I$ of the object beam, the LO beam frequency $\omega_{LO}$ being kept constant. The four phases  hologram $H$ is calculated and the  +1 grating order signal  $ E E_{LO}^*$ is selected in the Fourier space \cite{cuche2000spatial}.

For each signal frequency $\omega$, a sequence of frames $I_n$ is  recorded, and  four phases holograms $H$ and $\tilde H$ are calculated in real and Fourier space:
\begin{eqnarray}\label{Eq_H_4_phases_bis}
    H(x,y)&=&(I_1(x,y) -I_3(x,y) \\
 \nonumber    &&~~~~~+j(I_2(x,y)-I_4(x,y))\\
\nonumber  {\tilde H}(k_x,k_y)  &=& \textrm{FFT} \left[ H(x,y) \right]
\end{eqnarray}
where $\textrm{FFT}$ is the 2D Fourier transform operator.

Figure \ref{Fig_Fig_accurate_phase} (a) displays  $|\tilde H|^2$ that is obtained with $\omega_{AOM2} - \omega_{AOM1} = +\omega_{CCD}/4$.  The brighter point on the left (within circle 1) corresponds to the +1 grating order signal  $|E E_{LO}^*|^2$, while the  point on the right (within circle 2) is the -1 grating order signal $|E^* E_{LO}|^2$, whose intensity is much lower. Here in Fig. \ref{Fig_Fig_accurate_phase} (a), the 4 phase detection made with  $\omega_{LO}-\omega= +\omega_{CCD}/4$,
allows one to select the +1 grating order and to
 reject the -1 order.  In figure \ref{Fig_Fig_accurate_phase} (b), in contrast,
with $\omega_{LO}-\omega= -\omega_{CCD}/4$ one selects the -1 grating order and rejects the +1 order, since the -1  image is bright, while
the +1 image  is dark.

We have measured the weight of the $\pm 1$ signals $W_{\pm 1}$  defined by
\begin{eqnarray}\label{Eq_W}
    W_{\pm 1}&=& \sum_{{\pm 1}}   |{\tilde H}(k_x,k_y)|^2
\end{eqnarray}
where $\sum_{{\pm 1}}$ is the sum over a $10\times 10$ pixel region centered
on the $\pm 1$  peaks.
By sweeping the AOM1  frequency, we have studied how $W_{\pm 1}$  varies with $\omega$. Figure \ref{Fig_Fig_accurate_phase} (c,d) shows $W_{\pm 1}$ as
a function of $\omega_{LO}-\omega$.
When the four-phase condition is
fulfilled, i.e., when $\omega_{LO}-\omega=+\omega_{CCD}/4$, the weight $W_{+ 1}$ of +1
signal  is maximum (i.e. $W_{+ 1} \sim 1$) and   the twin signal
is nearly zero (i.e. $W_{- 1} \sim 10^{-4}$). Conversely, when $\omega_{LO}-\omega=-\omega_{CCD}/4$, $W_{-1}$ is large and $W_{+1}$ nearly zero.

The weight $W_{+ 1}$  of   $|E E^*_{LO}|^2$ term is proportional to $|\eta|^2$, where $\eta$ is the $E E^*_{LO}$ detection efficiency. For four phases detection with $4N$ frames i.e. for
\begin{eqnarray}\label{Eq_4N_phases}
 H&=& (1/4N) \sum_{n=1}^{4N} j^{n-1}I_n
\end{eqnarray}
we have \cite{verpillat2010digital}:
\begin{eqnarray}\label{EQ__eta}
  \eta(x)&=& \frac{1}{4N T} \sum_{n=0}^{4N-1} j^n \int_{t=n T_{\textrm{CCD}}-T/2}^{n T_{\textrm{CCD}}+T/2} e^{2\pi x t} dt\\
\nonumber    &=& \textrm{sinc}(\pi x T) \times \frac{1}{4N} \sum_{n=0}^{4N-1} j^n e^{2\pi n x T_{\textrm{CCD}}}
\end{eqnarray}
where $x= (\omega-\omega_{LO}-)/2\pi$ is the heterodyne beat frequency in Hz, $T$ the frame exposure time and $T_{CCD}=2\pi/\omega_{CCD}$ the frame to frame time.

In Eq. \ref{EQ__eta}, the factor $\textrm{sinc}(\pi x T)$, which has been already introduced in Eq. \ref{Eq_sinc_time}, corresponds to the integration
of the beat signal, whose frequency is non-zero, over
the camera exposure time $T$. Because of Eq. \ref{Eq_4N_phases}, the summation over the $4N$
frames of Eq. \ref{EQ__eta} is made with a  phase
factor $j^n$. To the end, the factor $1/4N$ is a normalization
factor that is the inverse of the number of terms within the summation. With this normalization
factor   the maximum of $|\eta(x)|$ is slightly lower than 1.

As depicted in Fig. \ref{Fig_Fig_accurate_phase} (c), the experimental results for
$W_{\pm1}$ (points) agree with $|\eta|^2$, where $\eta$ is given by  Eq. \ref{EQ__eta} with $T=T_{CCD}=0.1$ s and $4N=4$ (solid gray
curves). This means that heterodyne holography is  able to perform phase shifting with very accurate phase  \cite{atlan2007accurate}.  The hologram $H$ can thus be calculated without having to take account for the phase errors  of the experimental setup.


We must notice also that  the $+1$ grating order  image corresponds to the $\omega_{LO} +  \omega_{CCD}/4$ signal, while the  $-1$  image corresponds to $\omega_{LO} - \omega_{CCD}/4$.
If the frequency  spectrum of the signal field $E$ is broad and cover both $\omega_{LO} \pm  \omega_{CCD}/4$ frequency components, the $\pm 1$ images  both exist.  One must thus work off axis in order to separate them, and to avoid image alias.

The purpose of Fig. \ref{Fig_Fig_accurate_phase} experiment is to measure the frequency response  of the holographic detection. It turns out that the detection  bandwidth BW  (that is the width of the frequency response) is very narrow (about $\pm$ 2.5 Hz). It is nevertheless possible to detect signals with  Doppler shift or  Doppler broadening much larger than BW. Indeed, Fig. \ref{Fig_Fig_accurate_phase}  shows also that the detection  is made at a frequency close to $\omega_{LO}$ with an offset of + 2.5 Hz. By changing the frequency $\omega_{LO}$ it is thus possible to tune the holographic detection at any wanted frequency. In \cite{gross2003shot} for example, the signal is an ultrasonic sideband whose frequency $\omega$ is shifted by 2.2 MHz with respect to the object illumination frequency ($\omega \simeq \omega_I + 2.2$ MHz). This sideband signal has been detected by tuning $\omega_{LO}$ near $\omega$, the best detection efficiency being obtained with $\omega_{LO}= \omega + \omega_{CCD}/4$.

\subsection{Bandwidth BW of the holographic heterodyne detection}

\begin{figure}[h]
\begin{center}
  \includegraphics[width=7.5 cm]{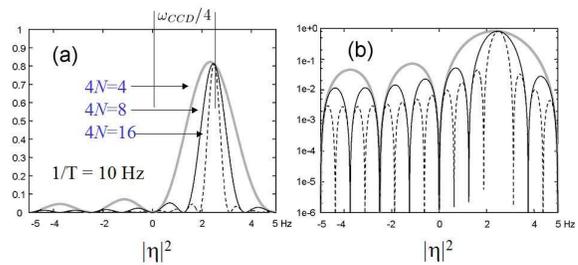}
  \caption{Detection efficiency  $|\eta|^2$   for four phases heterodyne detection with $4N$ frames: $4N=4$ (heavy grey line), $4N=8$  (solid black line), and $4N=16$ (dashed
black line). Calculation is done for $T=T_{\textrm{CCD}}=0.1$ s. Vertical axis axis is    $|\eta|^2$   in
linear (a) and logarithmic (b) scales. Horizontal axis is  $(\omega_{LO}-\omega)/2\pi$  in Hz}\label{Fig_transp12}
\end{center}
\end{figure}

The bandwidth BW of the holographic heterodyne detection  depends on the acquisition condition, in particular on the total exposure time. To illustrate this point, we have plotted, for four phases detection with $4N$ frames,  the detection efficiency $|\eta(x)|^2$ given by Eq. \ref{EQ__eta}.

As shown in Fig. \ref{Fig_transp12}, the detection bandwidth  BW  (i.e. the width of the frequency response $|\eta(x)|^2$) is inversely proportional to the exposure
time: $\textrm{BW} = (4T)^{-1}$,  $(8T)^{-1}$ and $(16T)^{-1}$  for 4, 8 and 16 frames respectively. This illustrates  the coherent character of the detection.

\subsection{Double filtering of the zero order signal and shot noise}

Since laser emission and photodetection on a  camera
pixel are random processes, the signal $I$ that is detected on each
pixel exhibits shot noise. This noise is gaussian and its standard deviation $\sigma$ is equal to $\sqrt{I}$, where $I$ is the pixel signal expressed in photo electron Units ($e$).
When performing heterodyne holography in dim light condition i.e. when $|E| \ll |E_{LO}|$ the detection sensitivity is generally limited by the shot noise on the local oscillator, since  $I \simeq |E_{LO}|^2 $.

To illustrate this point let us consider an example. Digital holography is made with an 8 bits camera, whose full well capacity is 2 10$^4$ $e$. The camera signal $I$ varies from 0 to 255 Digital Counts (DC) and the  camera "gain" is  $G = 78 $ $e$/DC. In a typical situation, the  local oscillator  is adjusted to half saturation, and the local oscillator is  $|E_{LO}|^2 = 10^4~\textrm{e} = 128 ~\textrm{DC}$.

The shot noise, whose   standard deviation is   $\sigma=\sqrt{I}=100$ $e$,  is much larger than the camera read noise ($\sigma<30$ $e$), and than the quantization noise of the camera Analog Digital converter ($\sigma= 78/\sqrt{12}=6.5$ $e$).
On the other hand, the shot noise equivalent signal is 1 $e$ per pixel. Indeed, if we consider  a very low signal: $|E|^2 = 1$ $e$, the +1 grating order holographic term    $|E E_{LO}^*|=100$ $e$ is equal to the shot noise variance $\sigma=100$ $e$.

This result, which  is valid for 1 frame, remains valid whatever the number of frames $4N$ is. Indeed, shot noise  is a broadband white noise. The noise that is detected is thus proportional to the product of
the total exposure time $4NT$  with the detection bandwidth BW, which is  proportional
to $(4NT)^{-1}$. The shot noise equivalent signal does  not depend on the number of frames $4N$. The   shot noise is  thus always equal to 1 $e$ per pixel.

\begin{figure}[]
\begin{center}
  \includegraphics[width=7.5 cm]{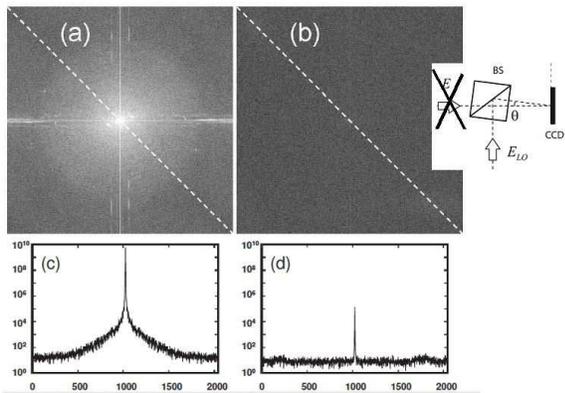}
  \caption{ (a,b) Images of the Fourier holograms  ${\tilde H}^2$  obtained without signal (i.e. with $|E|^2=0$). (c,d) Cut along the white dashed diagonal lines of the (a,b) images. Vertical axis is  ${\tilde H}^2$  along the cut averaged over 11 pixels:  ${\tilde H}^2$ is  plotted in log scale.  Horizontal axis is the pixel index (0...2047) along the $k_x$ and $k_y$ axis of the Fourier space. The $1280\times 1024$ recorded hologram  is padded into $2048\times 2048$ calculation grid. The Fourier space is thus  $2048\times 2048$ pixels, whose  size is  $2\pi /(2048 D_{pix})$, where $D_{pix}=6.7 \mu$m is the  camera pixel size.  Images and cut are calculated with one frame  (a,c) and   four frames holograms (b,d). See \cite{gross2008noise}. }\label{Fig_Fig_double_filtering}
\end{center}
\end{figure}

So, at this point a question arose: how it is possible to  reach this shot noise limit
in real time holographic experiment? As discussed  above,  the read noise and the   quantization noise are lower than shot noise. The last extra noise  that must be
considered is  the technical noise on the LO beam, because the LO signal is dominent since $|E_{LO}|^2\gg |E|^2$. This technical noise is the sum of the noises that differ from shot noise and that affect the LO beam signal. The  noise on the power supply of main laser,  the vibration of the setup, or the noise on the AOMs or on the RF signal that drives the AOMs contribute to technical noise. Whatever its origin, the LO technical noise is  highly correlated from pixel to the next, since the  LO beam that reaches the camera is  flat field ($E_{LO}|^2$ varies slowly with position $x$ and $y$).

It is thus possible to reach the shot noise limit by filtering  the LO technical noise.
A control holographic experiment made without signal (i.e. with $|E|^2=0$)  illustrates this filtering  \cite{gross2008noise}. Figure   \ref{Fig_Fig_double_filtering}(a, b) shows the  Fourier space holograms  $|{\tilde H}|^2$ that are obtained with one (a) and four frames (b).
With one frame (a) i.e. with $H=I_1$, the LO signal $|{\tilde H}|^2$, which is located in the center of the calculation grid, is  brighter and extends over a
larger area than with four frames  i.e. with $H=(I_1-I_3)+ j(I_2-I_4)$ (b).

To make a quantitative analysis of these images, we have plotted, on Figs. \ref{Fig_Fig_double_filtering} (c) and (d),    $|{\tilde H}(k_x, k_y)|^2$  along the diagonal dashed lines of Figs.  \ref{Fig_Fig_double_filtering} (a) and (b) respectively.
We have made diagonal cuts in order to explore
zones of the Fourier space that are far from  regions with $k_x$ or $k_y\simeq 0$ where the Fast Fourier Transform (FFT) alias can be observed as in Fig.  \ref{Fig_Fig_double_filtering}(a). For $k_x=k_y=0$, we obtain a peak on the
Figs. \ref{Fig_Fig_double_filtering} (c) and (d) cuts. These peaks corresponds to the
 LO field $|E_{LO}|^2$.

 In the one frame case (Fig. \ref{Fig_Fig_double_filtering} (c) ), the peak is  much larger and  much broader, than in the four frames case (Fig. \ref{Fig_Fig_double_filtering} (c) ). It results that most of the Fourier space is polluted by the
one frame LO parasitic signal, which  is several orders of
magnitude larger than its four frames counterpart. The one frame
LO parasitic signal is much larger than the shot-noise
limit, which is equal, within a few percent as experimentally verified, to the four frames  noise  floor seen on Fig. \ref{Fig_Fig_double_filtering} (d).

Here, in the  control holographic experiment made without signal, the shot-noise limit can be reached with four frames
detection, because  the LO signal  $|E_{LO}|^2$ (and the LO technical noise) are filtered off by a double filtering procedure in space and time.
\begin{itemize}
  \item Filtering in space is made by selecting, in the Fourier space, the off axis region where the +1 grating order signal $E E^*_{LO}$ is located. This region is far from the center of the Fourier space (see for  example Fig. \ref{Fig__124phase} (a) ), where the  LO beam signal $|E_{LO}|^2$ is located.
  \item  Filtering in time is made by the demodulation equation $H=(I_1-I_3)+j(I_2-I_4)$, which calculate $H$ from difference of frames recorded at different times. Since $|E_{LO}|^2$ does not vary with time,  $|E_{LO}|^2$ is  filter off by the demodulation equation, as seen on Eq. \ref{Eq_H_4_phases}.
\end{itemize}

\section{Heterodyne holography  examples of experiments}

To illustrate the advantages of heterodyne holography, we will present here two examples of experiment.

\subsection{Sideband holography }


The main advantage of heterodyne holography is its ability to detect the holographic signal at a frequency that is different from that of illumination. To illustrate this point,  we will first consider the detection of the light scattered by a vibrating object.

\subsubsection{Optical signal scattered by a vibrating object}

\begin{figure}[h]
\begin{center}
\includegraphics[width=6.5 cm]{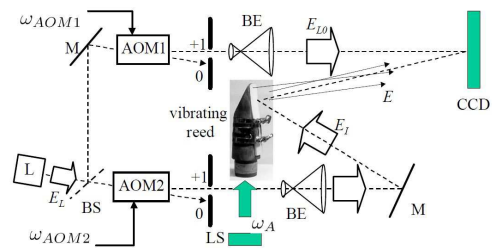}
 \caption{Heterodyne holography setup applied to analyse vibration of a clarinet reed. L: main laser; AOM1, AOM2: acousto-optic modulators; M: mirror; BS:
beam splitter; BE: beam expander; CCD:  camera; LS: loud-speaker exiting the vibrating  clarinet
reed at frequency $\omega_A/2\pi$.}\label{Fig_fig_setup_reed}
\end{center}
\end{figure}

   Figure \ref{Fig_fig_setup_reed} shows a typical example of vibration holographic experiment setup in reflection geometry. The object is a clarinet reed, whose vibration is excited by a loudspeaker LS. The vibration is studied by heterodyne holography  with a reference local oscillator beam (field $E_{LO}$) that can be frequency shifted with respect to the  illumination beam  (field $E_I$).

Let us consider that the  objet  (the clarinet reed) that vibrates   at frequency
$\omega_A$  with amplitude $z_{max}$. The displacement $z$ along the out of plane direction is
\begin{equation}\label{Eq_z_t}
  z(t) = z_{max} \sin{\omega_A t}
\end{equation}
In backscattering geometry, this corresponds to a
phase modulation $\varphi(t)$ of the signal:
\begin{eqnarray}\label{Eq_varphi_t}
 \varphi(t) &=& 4\pi z(t)/\lambda\\
  \nonumber  &=&\Phi \sin{\omega_A t}
\end{eqnarray}
where $\lambda$ is the optical wavelength and  $\Phi$ the amplitude of the phase modulation of the signal  at angular frequency $\omega_A$:
\begin{eqnarray}\label{Eq_Phi}
\Phi&=&4\pi z_{max}/\lambda
\end{eqnarray}
Let us define the slowly varying complex  amplitude $E(t)$ of the field ${\cal E}(t)$ scattered by the vibrating object. We have:
\begin{eqnarray}
{\cal E}(t)&=& E(t) e^{j\omega_I t} + \textrm{c.c.}
\end{eqnarray}
Because of the Jacobi-Anger expansion, we get:
\begin{eqnarray}\label{Eq_cal_E_sum_over_harmonic}
\nonumber E(t)&=&E ~e^{j\varphi(t)}=E ~e^{j \Phi \sin{\omega_A t}  }\\
\nonumber &=& E ~\sum_m J_m(\Phi)~e^{j n\omega_A t}
\end{eqnarray}
where $E$ is the complex amplitude without vibration, and $J_m$
 the mth-order Bessel function of the first kind, with
$J_{-m}(z)=-1^m J_m(z)$ for integer $ m$ and real $z$.
The scattered
field ${\cal E}(t)$ is then the sum of the carrier and sideband field components ${\cal E}_m(t)$ of frequency $\omega_m$, where $m$ is the sideband index with:
\begin{eqnarray}\label{Eq_cal_E_m}
 {\cal E}(t) &=& \sum_{m=-\infty}^{+\infty}{\cal E}_m(t)\\
\nonumber   {\cal E}_m(t) &=& E_m e^{j\omega_m t} +  E_m^* e^{-j\omega_m t}\\
\nonumber   \omega_m &=& \omega_I + m \omega_A
\end{eqnarray}
where $ E_m$ is the complex amplitude of the field component ${\cal E}_m(t)$. Note that $\omega_0=\omega_I$ is the  illumination optical frequency. Equation \ref{Eq_cal_E_sum_over_harmonic} yields:
\begin{equation}\label{Eq_E_m}
    E_m = J_m(\Phi)~ E
\end{equation}

\begin{figure}
\begin{center}
\includegraphics[width=4 cm]{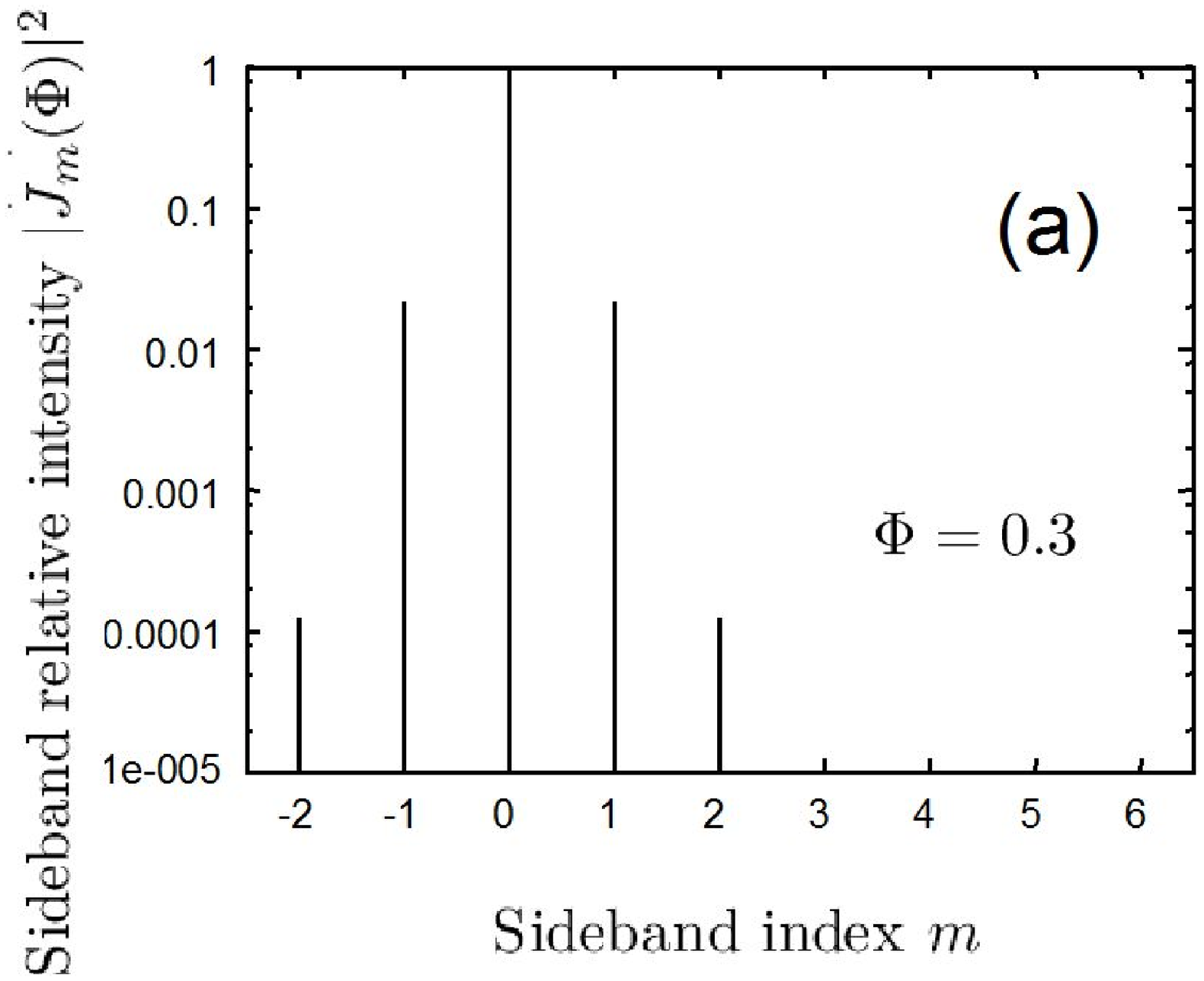}
\includegraphics[width=4 cm]{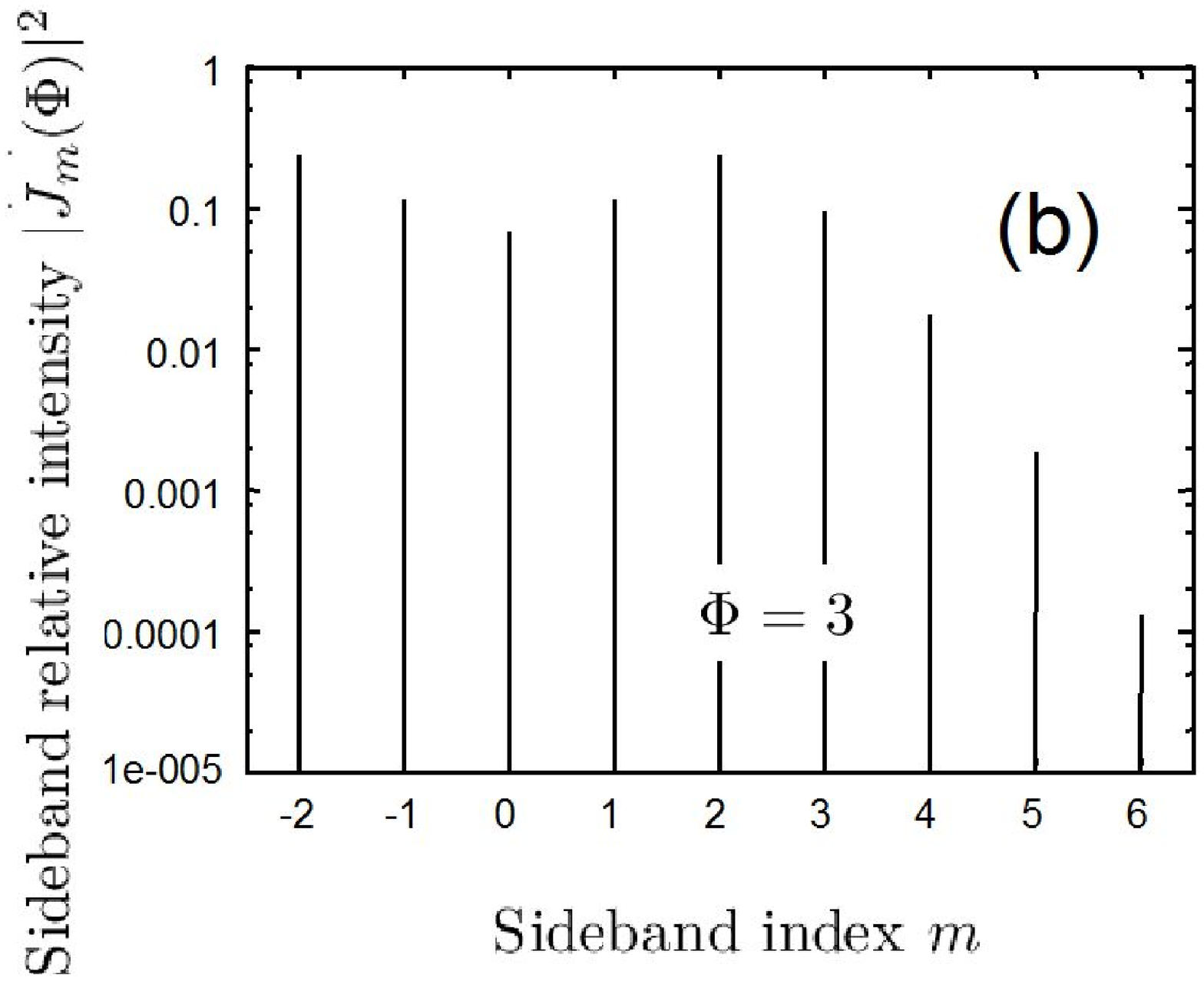}
 \caption{Relative amplitude $|E_m|^2/|E|2=|J_m(\Phi)|^2$ of the sideband component $m$ for an  amplitude modulation of the phase equal to  $\Phi=0.3$  (a) and $\Phi=3.0$ (b)  rad.}\label{Fig_fig_sideband _energy}
\end{center}
\end{figure}
Figure \ref{Fig_fig_sideband _energy} presents the distribution of the field energy   on the sidebands components $|E_m|^2$ for $\Phi=0.3$ and $\Phi=3$. If the amplitude of modulation $\Phi$ is low (see Fig. \ref{Fig_fig_sideband _energy} (a) ), most of the energy is on the carrier: $|E_0|/|E|^2\simeq 1$, and energy $|E_m|^2$  decreases rapidly with the sideband index $m$. If the amplitude $\Phi$ is large (see Fig. \ref{Fig_fig_sideband _energy} (b) ), the energy of the carrier is low: $|E_0|/|E|^2 \ll 1$, while  energy is distributed over many sidebands  $|E_m|^2$ .

%
%
%

\subsection{Selective detection of the sideband components $E_m$: by sideband holography  \cite{joud2009imaging}}
\label{section_Selective detection of the sideband components}

\begin{figure}
\begin{center}
\includegraphics[width=7.0 cm]{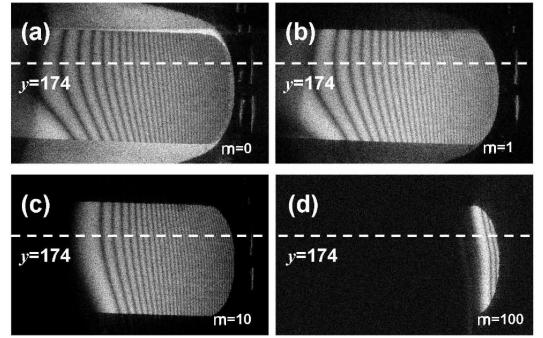}
 \caption{Reconstructed holographic images of a clarinet reed vibrating at frequency $\omega_A/2\pi =
2143$ Hz perpendicularly to the plane of the figure. Figure (a) shows the carrier image
obtained for $m = 0$. Fig. (b)-(d) show the frequency sideband images respectively for $m = 1$,
$m = 10$, and $m = 100$. A logarithmic grey scale has been used. See  \cite{joud2009imaging}. }\label{Fig_fig_clarinet_reed}
\end{center}
\end{figure}
Heterodyne holography 
%
is well suited to detect the vibration sideband components $E_m$.
To selectively detect  by  four phase demodulation the sideband $m$ of frequency $\omega_m$,
the frequency, $\omega_{LO}$ must be adjusted to fulfill the condition :
\begin{eqnarray}\label{Eq_Delata omega 4phase_sideband}
     \omega_{LO} &=& \omega_I + m \omega_A +\omega_{CCD}/4\\
 \nonumber    &=& \omega_m+\omega_{CCD}/4\\
 \nonumber  \textrm{with} ~~\omega_m&=&\omega_I + m \omega_A
\end{eqnarray}

Figure \ref{Fig_fig_clarinet_reed} shows  images obtained by detecting different sideband $m$ of a clarinet reed \cite{joud2009imaging}.
The clarinet reed is
attached to a clarinet mouthpiece and its vibration is driven by a sound wave propagating inside
the mouthpiece, as in playing conditions, but the sound wave is created by a
loudspeaker excited at frequency $\omega_A$ and has a lower intensity than inside a clarinet. The excitation frequency
is adjusted to be resonant with the first flexion mode (2143 Hz) of the reed.

Figure \ref{Fig_fig_clarinet_reed} (a) is obtained at the unshifted carrier frequency $\omega_0$. It corresponds to an image obtained by time averaging holography \cite{picart2003time}.
The left side of the reed
is attached to the mouthpiece, and the amplitude of vibration is larger at the tip of the reed
on the right side; in this region the fringes become closer and closer and difficult to count.
The mouthpiece is also visible, but without fringes since it does not vibrate.

Similar images
of vibrating objects  have been obtained in \cite{demoli2004detection,Demoli_2005,picart2005some,picart20052d,leval2005full,Picart_2010_study}, with more conventional techniques performing the holographic detection at the illumination frequency (i.e. carrier frequency). The quality of the images obtained in \cite{demoli2004detection,Demoli_2005,picart2005some,picart20052d,leval2005full,Picart_2010_study} are generally lower. Indeed, single frame holography is used in \cite{demoli2004detection,Demoli_2005,picart2005some,picart20052d,leval2005full,Picart_2010_study}.    The total amount of signal is thus lower. Moreover, single frame holography is unable to filter off the LO beam noise by making difference of images. The technical noise  and the order zero alias are much bigger as discussed in \cite{gross2008noise}.

Figures \ref{Fig_fig_clarinet_reed} (b) to (d) show images obtained for the sidebands $m=1$, 10 and 100. As
expected, the non-vibrating mouthpiece is no longer visible. Figure \ref{Fig_fig_clarinet_reed} (b) shows the $m = 1$
sideband image, with $J_1$ fringes that are slightly shifted with respect to those of $J_0$. Figure
\ref{Fig_fig_clarinet_reed} (c) shows the image of sideband  $m = 10$ and Fig.
\ref{Fig_fig_clarinet_reed} (d) the  $m = 100$ one. The left side region of the image
remains dark because, in that region, the vibration amplitude is not sufficient to generate these
sidebands, $J_m(z)$ being evanescent for $z < m$.

\begin{figure}
\begin{center}
\includegraphics[width=9 cm]{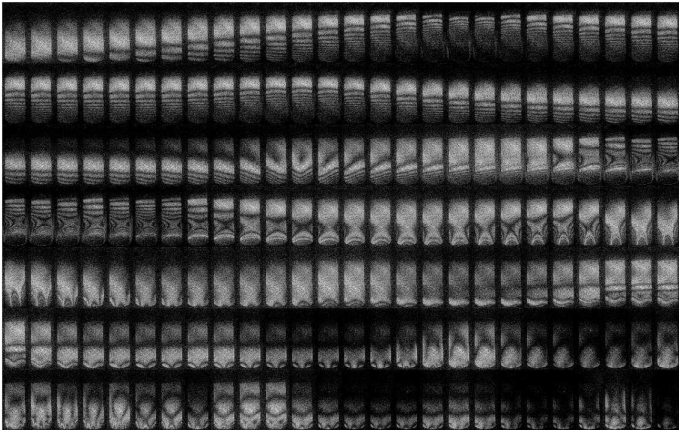}
 \caption{Clarinet reed  reconstructed images  obtained on sideband $m=1$.  Frequency $\omega_A$ is swept  from 1.4 KHz up to 20 KHz, and images are displayed from left to right and   top to bottom ($26 \times 7$ images). The display is made with arbitrary grey scale for the intensity $|E_0(x,y)|^2$. See  \cite{taillard2014statistical}. }\label{Fig_fig_taillard}
\end{center}
\end{figure}


In a typical heterodyne holography setup, RF generators   drive  AOM1 and AOM2 at $\omega_{AOM1}$ and  $\omega_{AOM2}$,
the camera at  $\omega_{CCD}$, and the vibration frequency at $\omega_{A}$. These genetators use a common 10 MHz reference frequency, and  are driven by the computer.   It is then possible to automatically  sweep $\omega_{A}$, $\omega_{AOM1}$  and  $\omega_{AOM2}$ in order to fulfil Eq. \ref{Eq_Delata omega 4phase_sideband} so that detection remains ever tuned on a given sideband.  Figure \ref{Fig_fig_taillard} shows an example \cite{taillard2014statistical}. A series of $26 \times 7$ images of a clarinet reed  are obtained on sideband $m=1$  by sweeping the frequency $\omega_A$ from 1.4 kHz up to 20 kHz by steps of 25 per cents (factor $1$, $1.25$, $(1.25)^2$, $(1.25)^3...$). The amplitude of the excitation signal is  exponentially increased in the range 1.4 to 4 kHz, from 0.5 to 16 V, then kept constant at 16 V up to 20 kHz. This crescendo limits the amplitude of vibration of the first two resonances of the reed.  The different vibration  modes  of the reed can be easily recognized on the reconstructed reed images of Fig. \ref{Fig_fig_taillard}.

Sideband heterodyne holography is a very powerful technique, which has been used to image objects whose vibration amplitude are both large  \cite{joud2009fringe} and small \cite{psota2012measurement,psota2012comparison,verrier2013absolute,bruno2014holographic,psota2014high}. Combined with stroboscopic illumination/detection \cite{verpillat2012imaging} or with sideband correlation analysis \cite{verrier2015vibration}, sideband heterodyne  holography has been made  sensitive to the phase of the  vibration.

\subsection{Analyse of Doppler spectrum  of light that travel through the breast \cite{gross2005heterodyne}}

\begin{figure}[h]
\begin{center}
  \includegraphics[width=7.5 cm]{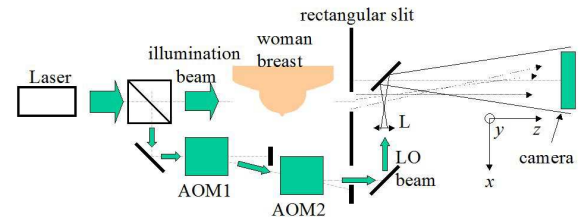}
  \caption{Doppler spectrum experimental setup. BS: beam splitter; AOM1, AOM2: acousto optic modulator;  L: short focal lens. Laser: near infrared laser (70 mW, 780 nm). }\label{Fig_Fig_transp_27_setup}
\end{center}
\end{figure}

The purpose of the second example of experiment is to measure the spectral broadening of the light that travel through a woman breast in vivo. This experiment has been made in the context of ultrasonic photon tagging \cite{leutz1995ultrasonic,wang1995continuous,leveque1999ultrasonic,murray2004detection,ramaz2004photorefractive}, whose purpose is to detect breast cancer. In that context the measure of the light spectral broadening, which is related to the breast transmitted light correlation time is important, since light decorrelation is important limitation of the performance of the ultrasonic photon tagging method.

Figure \ref{Fig_Fig_transp_27_setup} shows the heterodyne holography experimental setup. The object, whose image is reconstructed by holography,  is a  rectangular slit that is back illuminated by laser beam (70 mW, 780 nm) that travel through the breast.
%
%
The position of the lens L is adjusted so that the LO beam reach the camera off axis with respect to the object (the slit). Moreover, the slit to camera distance is made equal to the radius of curvature of the LO beam that reach the camera. By this way, the reconstruction of the +1 and -1 grating order images of the slit are made by a simple Fourier transform.

\begin{figure}[h]
\begin{center}
  \includegraphics[width=7.5 cm]{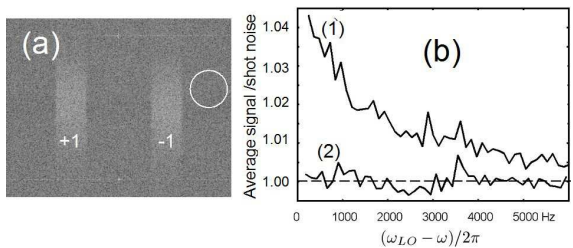}
  \caption{(a) Four frames Fourier space hologram $|{\tilde H}|^2$  obtained in the breast experiment  for $\omega_{LO}=\omega_I$. Display is made in arbitrary logarithmic gray scale. (b) Ratio of $|{\tilde H}|^2$  averaged within the +1 and -1 images of the slot,  versus $|{\tilde H}|^2$  averaged within the white circle quite zone, as a function of $(\omega_{LO}-\omega_I)/2\pi$ (curve 1) and $(\omega_{LO}-\omega_I)/2\pi + 100$ kHz (curve 2). }\label{Fig_fig_transp_27_imag_curve}
\end{center}
\end{figure}

Figure \ref{Fig_fig_transp_27_imag_curve} (a) shows the four frames Fourier space hologram $|{\tilde H}|^2$  that was obtained for $\omega_{AOM1}=\omega_{AOM2}$ i.e. for $\omega_{LO}=\omega_I$. The two brighter vertical bands are the  +1 and -1 grating order images of the slit, which are both on focus because of the peculiar location of lens L. The  +1 image
corresponds to the $|E|^2$ signal at frequency   $\omega_{LO}+\omega_{CCD}/4$, while the   -1 image
corresponds to  $\omega_{LO}-\omega_{CCD}/4$.

Because the breast inner motions are random, the frequency spectrum of the light scattered by breast is continuously broadened. In that spectrum the 4 phases detection that is made here selects the signal at  frequency   $\omega_{LO}+\omega_{CCD}/4$ within the +1 image, and the signal at  $\omega_{LO}-\omega_{CCD}/4$ within the -1 image. Both signals are non zero because the spectrum is continuous. Note that it is also possible to analyse the same data by making 2 phases detection with $H=I_1-I_2$. In that case, $H$ is a real quantity and the +1 and -1 images are both present too.

The detected signal  (i.e. the brightness of slit images)   is low for the following reasons.
\begin{itemize}
  \item Most of the incoming illumination light is back reflected because the breast is highly diffusing.
  \item The transmitted light is scattered in all directions and few light reach the camera.
  \item The light that reach the camera is Doppler broaden by the breast inner motions. Since this Doppler broadening is large with respect to the heterodyne detection bandwidth $\textrm{BW} = (4NT)^{-1} $, most of the light that reach the camera is not detected.
\end{itemize}

To explore the Doppler profile of the scattered photons,
the frequency of the LO beam offset $(\omega_{LO}-\omega_I)/2\pi$,  was
swept from 0 Hz up to 5000 Hz. To obtain a better signal, the 4 frames holograms  $|{\tilde H}|^2$ have  been averaged over 32 sequences of four frames. We have then measured the averaged signal within both the +1 and -1 images of the slit, and compared  this signal  with the averaged signal in a quite zone of the Fourier space, like the zone within the white circle in the right  of Fig. \ref{Fig_fig_transp_27_imag_curve} (a). The signal, within the quite zone,   corresponds to shot noise i.e. to 1 $e$ per pixel.

We have plot, on Fig. \ref{Fig_fig_transp_27_imag_curve} (b) (curve 1), the ratio of the  averaged +1 and -1 signal $|{\tilde H}|^2$ , versus the averaged  signal within the quite zone,  as a function of $(\omega_{LO}-\omega_I)/2\pi$ (curve 1). Because of the ratio, the vertical scale is the averaged signal $|E|^2$ in photo electron Units. In order to better visualise the background noise, we have also plotted on Fig. \ref{Fig_fig_transp_27_imag_curve} (b) (curve 2), the ratio versus $(\omega_{LO}-\omega_I)/2\pi + 100 $ KHz. Since 100 KHz is much larger than the half width of Doppler spectrum (about $ 1.5$ kHz), curve 2 corresponds to the background noise, that is effectively equal to   shot noise, i.e., 1 $e$.

This last experiment shows how it is possible to explore a Doppler spectrum by heterodyne holography. It illustrates also the sensitivity of Doppler holography, since the signal that is analyzed is very low: $|E|^2$ very less than 0.05 $e$ at maximum.

The Doppler heterodyne holography method, that is illustrated here, has been used in many other contexts. The method has been used  to detect  the tagged photons  in ultrasonic photon tagging experiments \cite{gross2003shot}, to study   brownian motion effects in coherent back scattering \cite{lesaffre2006effect}, and  to image    blood flow,  in mouse crania \cite{atlan2006frequency,atlan2007cortical},  rat eye  \cite{simonutti2010holographic} and fish embryo \cite{verrier2014laser}.

\section{Conclusion  }

In this paper, we have presented the digital heterodyne holography technique that
is able to fully control the amplitude, phase and frequency of both illumination and reference
beams. This control is made by acousto optic modulator, which are driven electronically by  RF generators.  Heterodyne holography is an extremely versatile and powerful tool,  able to perform automatic data acquisition in any holographic configuration with accurate  phase and shot noise sensitivity.

Heterodyne holography  is also able to perform the holographic detection at frequencies that are  different than the illumination one. This capability is especially useful for vibration  analysis, and for laser Doppler full field measurement.

We acknowledge
 ANR Blanc Simi 10 (n 11 BS10 015 02) grant,
ANR Blanc Simi 4 (n. 11-BS04 017 04) and
 and Labex Numev (convention ANR-10-LABX-20) grant for funding.

\bibliographystyle{unsrt}

\end{document}